\documentclass[floatfix,prl,superscriptaddress,twocolumn,letterpaper,nobalancelastpage]{revtex4-2}

\usepackage{mycommands}
\usepackage[T1]{fontenc}
\usepackage{url}

\begin{document}

\title{Spontaneous Circular Dichroism and Memory Effect from Surface Defects in Altermagnetic MnTe}
\author{Benjamin F. Mead}
    \affiliation{Department of Physics and Astronomy, University of Pennsylvania, Philadelphia, Pennsylvania 19104, USA}
    
    \author{Isaiah Gray}
    \affiliation{Department of Physics and Astronomy, University of Pennsylvania, Philadelphia, Pennsylvania 19104, USA}
    \affiliation{Present Address: Research and Exploratory Development Department,  Johns Hopkins University Applied Physics Laboratory, Laurel,  MD 20723}

    \author{Seng Huat Lee}
      \affiliation{Department of Physics, The Pennsylvania State University, University Park, PA 16802, USA}
      
    \author{Michael Chilcote}
    \affiliation{Materials Science and Technology Division, Oak Ridge National Laboratory,  Oak Ridge, TN 37831, USA}

   \author{Zi-Jie Yan}
      \affiliation{Department of Physics, The Pennsylvania State University, University Park, PA 16802, USA}

\author{Sijie Xu}
\affiliation{Department of Physics and Astronomy, Rice University, Houston, TX 77005, USA}
    \affiliation{Rice Laboratory for Emergent Magnetic Materials and Smalley-Curl Institute, Rice University, Houston, TX 77005, USA}

\author{Aparna Jayakumar}
\affiliation{Department of Physics and Astronomy, Rice University, Houston, TX 77005, USA}
    \affiliation{Rice Laboratory for Emergent Magnetic Materials and Smalley-Curl Institute, Rice University, Houston, TX 77005, USA}

 \author{Emilia Morosan}
    \affiliation{Department of Physics and Astronomy, Rice University, Houston, TX 77005, USA}
    \affiliation{Rice Laboratory for Emergent Magnetic Materials and Smalley-Curl Institute, Rice University, Houston, TX 77005, USA}

     \author{J\"orn W. F. Venderbos}
    \affiliation{Department of Physics, Drexel University, Philadelphia, Pennsylvania 19104, USA}
    \affiliation{Department of Materials Science and Engineering, Drexel University, Philadelphia, PA 19104, USA}

        \author{Cui-Zu Chang}
    \affiliation{Department of Physics, The Pennsylvania State University, University Park, PA 16802, USA}
    
    \author{Zhiqiang Mao}
    \affiliation{Department of Physics, The Pennsylvania State University, University Park, PA 16802, USA}

    \author{Pengcheng Dai}
    \affiliation{Department of Physics and Astronomy, Rice University, Houston, TX 77005, USA}
    \affiliation{Rice Laboratory for Emergent Magnetic Materials and Smalley-Curl Institute, Rice University, Houston, TX 77005, USA}
    
    \author{Matthew Brahlek}
    \affiliation{Materials Science and Technology Division, Oak Ridge National Laboratory,  Oak Ridge, TN 37831, USA}

    \author{Liang Wu}
    \email{liangwu@upenn.edu}
    \affiliation{Department of Physics and Astronomy, University of Pennsylvania, Philadelphia, Pennsylvania 19104, USA}

\date{\today}
\begin{abstract}
   Altermagnetism is a new class of magnetism with vanishing magnetization and macroscopic time reversal symmetry breaking. Hexagonal MnTe is a g-wave altermagnet with a 307~K N\'eel temperature, large spin splitting, and an anomalous Hall effect. However, the relationship between altermagnetism and the anomalous Hall effect in MnTe remains debated. Here, using optical magnetic circular dichroism, we observe hysteresis and a memory effect in a MnTe bulk crystal below 250 K. No hysteresis is seen after polishing the bulk crystal or in thin films. We conclude that the hysteresis in magnetic circular dichroism in the bulk crystal is from surface defects. The similar temperature dependence and a sign change  between circular dichroism and the anomalous Hall effect indicate that the origin of the anomalous Hall effect in transport is likely not intrinsic.

\end{abstract}
\maketitle

{\it Introduction.---}Altermagnetism (AM) is a recently identified class of compensated magnetic materials that host spin-dependent transport phenomena, typically associated with ferromagnets (FM), despite vanishing net magnetization, typically associated with antiferromagnets (AFM) \cite{Smejkalbeyond2022,Smejkalaltermag2022, Jungwirth2026signatures, Jungwirth2026AMspintronics, Song2025AMfunctional, Bai2024Frontiers, Tamang2025AFreview}. In AMs, two opposite spin sublattices are related by time reversal combined with a lattice rotation, rotoinversion, or mirror symmetry. This symmetry combination causes a nonrelativistic spin-splitting of the electronic bands with d-, g-, or i-wave symmetry, even in the absence of spin-orbit coupling. The spin-splitting is what enables AMs to exhibit spin-dependent transport phenomena including spin currents and an anomalous Hall effect (AHE) despite the nearly vanishing net magnetization. The AHE in altermagnets is particularly notable because it provides a route for electronic readout of time-reversed magnetic domains, directly addressing a significant obstacle for conventional AFM spintronics \cite{Jungwirth2026AMspintronics, Song2025AMfunctional, Jungwirthspintronics2016, Baltzspintronics2018}.

Hexagonal MnTe ($\alpha-$MnTe) has emerged as a textbook semiconducting altermagnet. $\alpha-$MnTe crystallizes in the NiAs-type structure (space group $P6_{3}/mmc$), with stacked triangular layers of Mn and Te atoms and a band gap of 1.4 eV. A magnetic transition occurs at $T_{N}=307$~K \cite{Szuszkiewiczneutron2005}. In the magnetic state each triangular Mn layer ferromagnetically orders with moments in the ab plane. Two adjacent layers show AFM-like anti-alignment forming the stacking unit. Nonrelativistic spin splitting (ARPES) \cite{KrempaskyARPES2024, Lee_APRES2024, Osumi_ARPES2024} and chiral magnons \cite{Liu_chiral2024,jost2025chiralaltermagnonmnte} have both been observed in $\alpha-$MnTe, providing the first direct evidence of AM order. The observation of the anomalous Hall effect (AHE) \cite{GonzalezBetancourt_AHE2023, Kluczyk_anomalous2024, Chilcote_stoichometry2024,liu2026strainAHE, smolenski2025strainberrycurvature, bey2024unexpectedtuninganomaloushall, zhou2026surfacestatedrivenanomaloushalleffect} provides indirect evidence of AM order, but is considered a proof of concept for electric read-out of 180 degree N\'eel vector switching. 

While the AHE is considered to arise from AM order, prior reports of the AHE in $\alpha-$MnTe show several
significant discrepancies between the observed AHE
and the AM order parameter. First, the AHE often turns on around 280 K, noticeably below the 307 K AM transition temperature \cite{GonzalezBetancourt_AHE2023, Chilcote_stoichometry2024, liu2026strainAHE, smolenski2025strainberrycurvature, bey2024unexpectedtuninganomaloushall, zhou2026surfacestatedrivenanomaloushalleffect}. Second, the hysteresis loop changes sign around 200 K with no shrinking of the coercive field. This sign change remains poorly understood, but is thought to be driven by changes in the Berry curvature of the electronic band structure instead of changes in the real-space-magnetization \cite{liu2026strainAHE, smolenski2025strainberrycurvature}. Lastly, the temperature dependence shows significant variation between samples \cite{GonzalezBetancourt_AHE2023, Kluczyk_anomalous2024, Chilcote_stoichometry2024,liu2026strainAHE, smolenski2025strainberrycurvature, bey2024unexpectedtuninganomaloushall, zhou2026surfacestatedrivenanomaloushalleffect} with Ref. \cite{liu2026strainAHE} reporting that the AHE occurs only when the thermal activation energy is 15--18~meV. These puzzling results have received different interpretations, raising questions about the relationship between the AHE and AM order in MnTe and motivating further investigation of how the magnetic state couples to the electronic transport properties.

\begin{figure*}
     \centering
     \includegraphics[width=\linewidth]{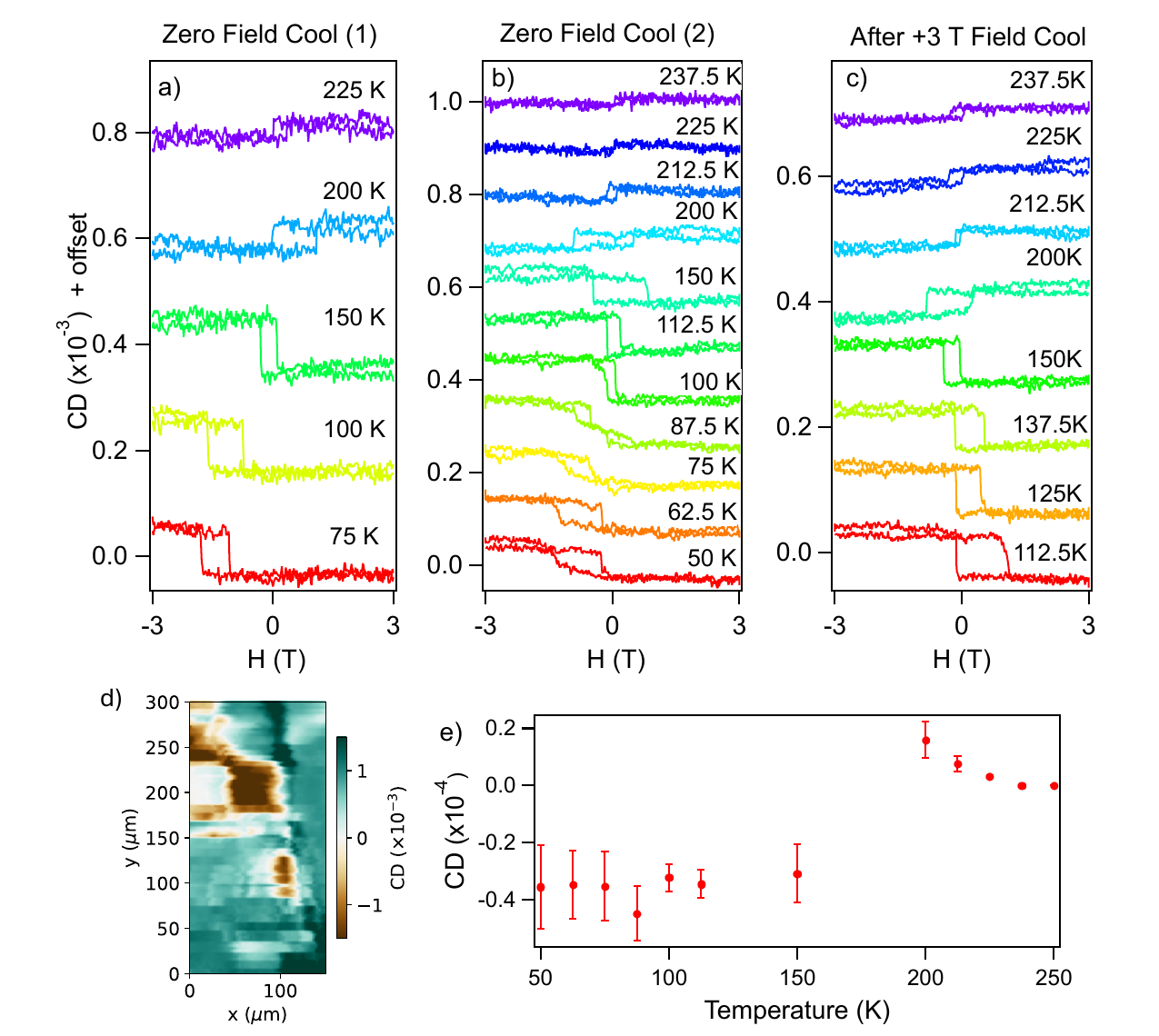}
     \caption{a, b) Temperature dependent circular dichroism hysteresis loops after consecutive zero field coolings from 300 K. Hysteresis turns on below 225--250 K and changes sign below 200 K. At lower temperatures, the hysteresis loop shifts to negative fields. c) Hysteresis loops after cooling the sample with a +3 T field from 300 K. The sign of the hysteresis does not change, but the loop now shifts to positive fields. Hysteresis loops in a-c were taken when warming after cooling the sample. No field was applied when changing temperature. d) CD mapping of the bulk crystal taken at +1 T and at 200 K. Simultaneous red and blue regions demonstrates time-reversed domains are simultaneously present. e) The anomalous component of the CD loops after zero field cooling as a function of temperature. Values are extracted from the data in Fig. 1b. The anomalous component is taken at $B_{c}$ the field the hysteresis loop is centered on. The anomalous component is defined as half the difference between the two domains at this field $(CD(B_{C}^{+})-CD(B_{C}^{-}))/2$ where $\pm$ refers to when sweeping from positive field $(+)$ and negative $(-)$ to $B_{c}$.
}
     \label{fig:temp}
\end{figure*}

Optical probes, such as magnetic circular dichroism (MCD), can study magnetic domains with micrometer spatial resolution \cite{Willard2004Microstructure, wu_magnetooptical_2024, Higo2018octupole, han2025largetopologicalmagneticoptical, Ni2021imagingAFMdomains, Ni2021strainimaging}. MCD is the difference in absorption of left-hand (LH) and right-hand (RH) circularly polarized light. MCD is proportional to the index of refraction difference for RH and LH circularly polarized light, $\triangle N=\frac{i\epsilon_{xy}}{\sqrt{\epsilon_{xx}}}$ where $\epsilon_{xy}$ and $\epsilon_{xx}$ are the transverse and longitudinal dielectric tensor components at a given photon energy \cite{YokoyamaMCD2008, Starketenor2020}. The dielectric constants are related to the optical conductivity. MCD probes the off-diagonal elements of the conductivity tensor making MCD an optical analogue to the AHE \cite{NagaosaAHE2010}.  Both x-ray MCD and magnetic linear dichroism have been previously employed to image AM domains in $\alpha-$MnTe thin films \cite{AminXMCD2024, Hariki_XMCD2024}. A non-zero static magneto-optical Kerr effect (MOKE) was predicted in $\alpha-$MnTe \cite{MazinMnTepredictions_2023}, time-resolved magneto-optical effects were observed in MnTe thin films \cite{GraytrMOKE2024}, and static MOKE was recently reported at infrared wavelengths \cite{watanabe2026magneto, yang2026giant}, but the microscopic origin remained unclear. 

\begin{figure*}
    \centering
    \includegraphics[width=\linewidth]{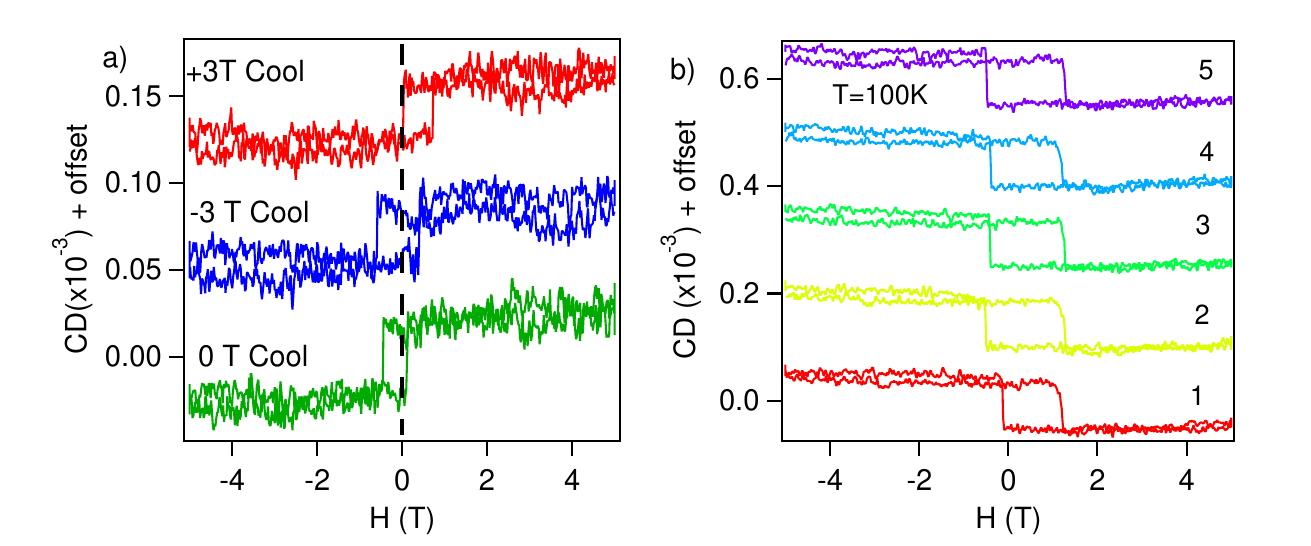}
    \caption{a) Field sweeps at 200 K after cooling the sample with $\pm3$ T and 0 T cooling. Switching the sign of cooling-field switches the direction of the hysteresis loop shifts. b) Consecutive field sweeps at 100 K immediately after cooling the sample with a +5 T field from 250 K. The coercive field becomes slightly larger after the first hysteresis loop. 
}
    \label{fig:FC}
\end{figure*}

Here we observe optical MCD hysteresis in a MnTe bulk crystal. The spontaneous (0 T) MCD onsets around 250 K, well below the 307 K N\'eel temperature. The observed hysteresis is shifted away from 0 T. Field cooling the sample shifts hysteresis in the direction of the applied field. After polishing the sample, no hysteresis or spontaneous MCD was observed. Thin films grown using  molecular beam epitaxy (MBE) on InP (111) also showed no hysteresis. We attribute the hysteresis to a surface-defect state and the shift to a magnetic memory effect. Our results suggest the MCD in MnTe originates from surface defects rather than bulk altermagnetic order.  
           
{\it Crystal growth.---}Our main MnTe bulk crystal was grown using a Bi-Te flux method. Under an argon atmosphere, Mn, Te, and Bi were mixed in a molar ratio of 1:16:10 and loaded into an alumina crucible. The crucible was then sealed in an evacuated quartz tube and heated to 900 $^{\circ}$C over 3 hours where it was held for 12 hours to promote homogeneous melting. Subsequently, the temperature was lowered to 610 $^{\circ}$C over 12 hours, followed by a slower cooling to 595 $^{\circ}$C over 200 hours to promote single-crystal growth. Finally, the Bi-Te flux was removed by centrifugation. Measurements were initially performed on shiny regions of the naturally formed facet without additional surface preparation. 

{\it Spontaneous magnetic circular dichroism.---}The field-dependent CD from this bulk crystal is shown in Fig. \ref{fig:temp}a. The CD was measured with a 633 nm continuous wave HeNe laser. Both the laser and applied magnetic field are normal to the sample surface. The sample was cooled from 300~K to the lowest temperature (50~K or 75~K) with no applied magnetic field and the hysteresis loop was measured. Then the sample was warmed with no applied field and the next loop was measured, until the highest temperature hysteresis loop was measured. Despite cooling with no magnetic field, the hysteresis loops are shifted and not centered at 0 T. At low temperatures the hysteresis loops are shifted towards negative fields. As the sample warms, the hysteresis loop reverses sign between 150~K and 200~K and becomes positively shifted. Hysteresis then disappears above 225~K. A second temperature sweep at the same spot is shown in Fig. \ref{fig:temp}b. The second run was taken after zero-field cooling the sample from 300 K then following the same warming and measurement process. Once again the hysteresis loop is shifted and changes sign between 150~K and 200~K. However, we observe that loops at the same temperature  (for example 100 K, 150 K, and 200 K) are different in these two runs. A mapping taken at 200 K is shown in Fig. \ref{fig:temp}d. The presence of red and blue regions, with opposite CD signs, shows time-reversal domains spontaneously form upon zero field cooling the sample.  

The temperature dependence of the spontaneous CD signal is shown in Fig. \ref{fig:temp}e. The spontaneous component is defined as half the difference between the two domains at the center of the hysteresis loop, $CD_{SP}=\frac{CD(B_{C}^{+})-CD(B_{C}^{-})}{2}$, where $B_{C}^{\pm}$ refers to the field halfway between the two coercive fields when swept from the maximum positive $(+)$ or negative field $(-)$. Below the hysteresis sign change, the spontaneous component, within experimental uncertainty, is nearly independent of temperature. This sign change is not readily explained by a direct correspondence between the spontaneous MCD and the AM order parameter. Nevertheless, the onset of hysteresis below the 307 K N\'eel temperature and the sign change around 200 K are consistent with prior AHE measurements on MnTe thin films and bulk crystals \cite{GonzalezBetancourt_AHE2023, Kluczyk_anomalous2024, Chilcote_stoichometry2024,liu2026strainAHE, smolenski2025strainberrycurvature, bey2024unexpectedtuninganomaloushall, zhou2026surfacestatedrivenanomaloushalleffect}. Based on the similarities in the temperature dependence, we hypothesize that the
MCD hysteresis shares a microscopic origin with the
AHE in MnTe.
\begin{figure*}
    \centering
    \includegraphics[width=\linewidth]{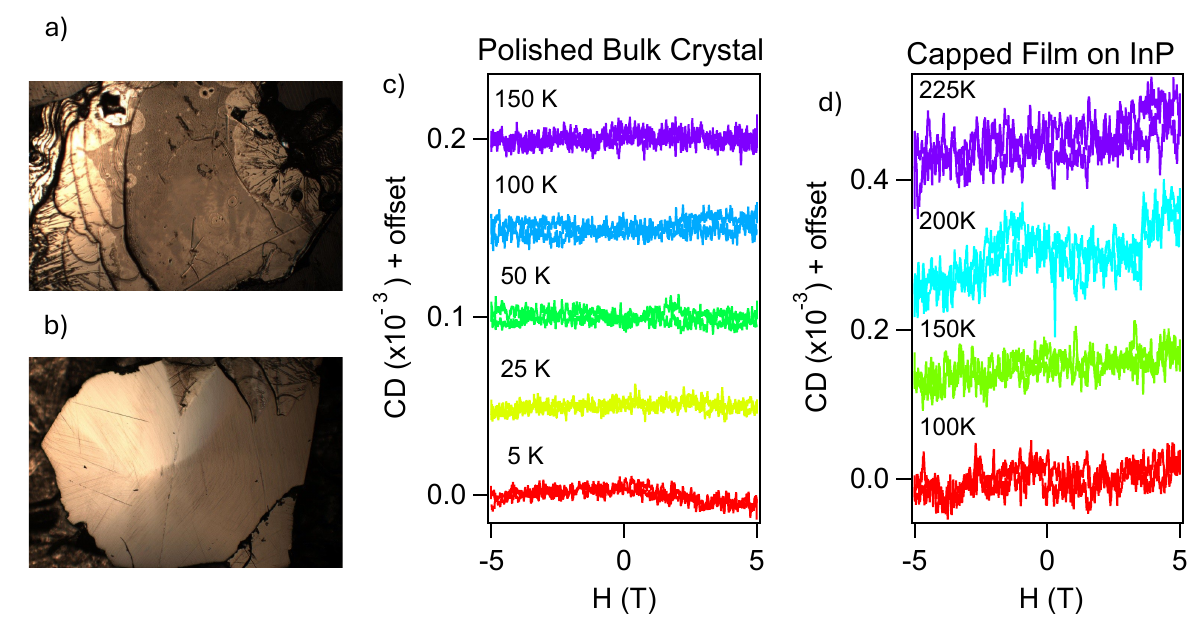}
    \caption{Microscope images of the MnTe bulk crystal with a 5x microscope objective a) before polishing and b) after polishing. Each image is about 3 mm by 3 mm. c) Hysteresis loops taken at select temperatures of a) polished MnTe bulk crystal and d) capped thin film on InP. Neither sample shows MCD hysteresis. The absence of hysteresis suggests the observed hysteresis above and exchange bias is sensitive to the surface of the bulk crystal.}
    \label{fig:films}
 \end{figure*}

To further probe the shift of the hysteresis, we field-cooled the crystal with +3 T and remeasured the hysteresis loops while warming the sample. As seen in Fig. \ref{fig:temp}c, field cooling the bulk crystal does not change the sign of hysteresis, but it shifts the hysteresis loop in the positive field direction. This is further shown in Fig. \ref{fig:FC}a. Hysteresis loops at 200 K were measured immediately after cooling the sample with the noted magnetic field. Switching the direction of the applied cooling field switches the direction in which the hysteresis loop is shifted. Fig. \ref{fig:FC}b shows five consecutive field sweeps at 100 K after field cooling the sample with +5 T. The hysteresis loop widens by 0.5~T after the first loop, then narrows by 0.1~T each on the third and fourth loops, and widens by 0.16~T on the fifth loop. These changes are larger than the approximate 0.02 T steps between data points, showing the hysteresis depends, in part, on the history of the applied magnetic field.

{\it Magnetic memory effect---.}A second magnetic system is required to pin domains and shift the hysteresis loop. The observed shift is reminiscent of exchange bias (EB) seen at FM-AFM interfaces, but occurs in a single material bulk crystal lacking a magnetic interface. Training effects in conventional EB can cause the shift to decrease, but wouldn't explain the change in coercive fields in Fig. \ref{fig:FC}b \cite{NOGUES1999exchange}. Training effects are typically absent when the AFM is a single crystal \cite{NOGUES1999exchange}. Since we observe one shifted hysteresis loop the second spin system, which pins the hysteresis loop, is not switched during our measurement and does not contribute to the observed MCD. If the second spin system was also switched, we expect the hysteresis loop to remain centered at zero field. This second system either has a spin configuration that does not couple to the applied magnetic field, or has a coercive field above $\pm5$ T. An alternative to interfacial EB is a dilute secondary spin population, distinct from the hysteresis causing system, shifts the hysteresis loop.

The secondary spin population hypothesis is motivated by prior studies on single material bulk crystals of the ferromagnetic Weyl semimetal Co$_{3}$Sn$_{2}$S$_{2}$. A field-tunable "positive exchange bias" in Co$_{3}$Sn$_{2}$S$_{2}$ was previously reported in  \cite{NoahtunableEB2022, Lachman2020exchangebias, Menilmemory2025} and recently attributed to a magnetic memory effect \cite{Menilmemory2025}. The memory effect caused the shape, saturation, and shift of the magnetization and AHE hysteresis loops to depend on temperature, maximum applied magnetic field, and the previously applied magnetic field. In addition to the 175 K Curie temperature \cite{VAQUEIRO2009neutron}, Co$_{3}$Sn$_{2}$S$_{2}$ was known to have another sample-dependent temperature scale of 125--150~K \cite{Zivkovic2022unraveling, Lachman2020exchangebias, Menilmemory2025, Lee2022} associated with a crossover from "bow tie" to rectangular hysteresis loops. At higher temperatures, the magnetization exhibits a sudden partial jump followed by a linear field dependence \cite{Zivkovic2022unraveling, Lachman2020exchangebias, Menilmemory2025}, consistent with a multidomain state stabilized by soft magnetic anisotropy. At lower temperatures, the magnetic anisotropy "hardens" and domain switching is complete as evidenced by rectangular hysteresis loops. If the secondary spin population does not flip, it pins the primary spins shifting the rectangular hysteresis loop away from 0 T. Ref. \cite{Menilmemory2025} proposes the memory effect arises from a dilute secondary spin population resides at either the sample surface or stoichiometric off-sites, but to our knowledge the microscopic origin remains unknown. A dilute secondary spin system with a large coercive field in our MnTe bulk crystal could explain the shifted hysteresis loop in the absence of a magnetic interface. Field cooling sets the secondary population's direction, thus determining the direction of the hysteresis shift.

Shifted hysteresis loops have been reported in dc transport on bulk crystals \cite{liu2026strainAHE} and thin films \cite{Chilcote_stoichometry2024} of MnTe. Ref. \cite{Chilcote_stoichometry2024} proposed a FM moment was pinned by bulk AM order causing EB in the magnetization curves. Unlike our bulk crystal, the EB in the thin films shifted hysteresis in the opposite direction as the cooling field. Ref. \cite{liu2026strainAHE} saw EB in strained bulk crystals which they attributed to pinning from impurities, but the EB was notable only below 75~K. To our knowledge, this is the first report of a shifted hysteresis loop in unstrained MnTe, and occurs at higher temperatures than in previous reports \cite{liu2026strainAHE, Chilcote_stoichometry2024}.

Evidence of multiple ferromagnetic systems has been in the AHE of MnTe thin films \cite{Chilcote_stoichometry2024, zhou2026surfacestatedrivenanomaloushalleffect}. Refs. \cite{Chilcote_stoichometry2024} and~\cite{zhou2026surfacestatedrivenanomaloushalleffect} both reported two jumps at two different coercive fields in the AHE in thin films grown with molecular beam epitaxy on InP and GaAs. Their explanations disagree. Ref. \cite{zhou2026surfacestatedrivenanomaloushalleffect} interpreted these loops as being from the top and bottom surface of their film while Ref. \cite{Chilcote_stoichometry2024} proposed an evolution of the FM moment from excess Mn. We note neither study reported MCD hysteresis and proposed it lies below the optical sensitivity limit. The sample to sample variation previously reported \cite{GonzalezBetancourt_AHE2023, Kluczyk_anomalous2024, Chilcote_stoichometry2024,liu2026strainAHE, smolenski2025strainberrycurvature, bey2024unexpectedtuninganomaloushall, zhou2026surfacestatedrivenanomaloushalleffect}, and our work, indicate defect-sensitive hysteresis loops in MnTe.

{\it Surface origin of hysteresis.---}To better probe the connection between bulk AM order and the observed hysteresis, we polished the MnTe bulk crystal and repeated the MCD measurements. Polishing the sample should not remove a bulk AM contribution as demonstrated by MOKE measurements of hematite Fe$_{2}$O$_{3}$ \cite{HaolinMOKE2026, LuoMOKE2026, yoshimochiMOKE2026}. Polishing the bulk crystal removes micrometers from the sample surface, much greater than the optical penetration depth typically around 50 nm for doped semiconductors above the gap, allowing us to probe freshly exposed crystallographic planes. As shown in Fig. \ref{fig:films}c all hysteresis disappears after polishing the bulk crystal. Figure \ref{fig:films} d shows no MCD hysteresis in a MBE-grown thin film on InP, despite showing an AHE in dc transport \cite{Chilcote_stoichometry2024}. The thin film was grown using the procedure described in Refs. \cite{Chilcote_stoichometry2024} and \cite{Chen2026PhaseSelection}. This is not the first report of null magnetic MCD in thin films with an AHE \cite{Chilcote_stoichometry2024, zhou2026surfacestatedrivenanomaloushalleffect} We also measured two separate bulk crystals grown using chemical vapor transport as described in Ref \cite{liu2026strainAHE}. Both additional bulk crystals showed an AHE in dc transport, but no spontaneous MCD. The microscopic mechanism for the AHE (and MCD) in MnTe may lie below the optical sensitivity limit in these samples giving an apparent absence of spontaneous MCD. The disappearance of spontaneous MCD hysteresis upon polishing and its absence in samples that exhibit an AHE indicate a non-robust surface effect that we propose arises from surface defects. 

If the spontaneous MCD and the AHE originate from impurities then a lower impurity concentration would reduce the spontaneous MCD below our optical sensitivity limit. Defect density is unlikely to be the only relevant quantity. The narrow thermal activation energy window to observe an AHE, reported in Ref. \cite{liu2026strainAHE}, is very consistent with disorder-driven variable-range hopping. Ref. \cite{liu2026strainAHE} also found that below 80 K, the scaling exponent of the anomalous Hall conductivity was in the localized hopping regime, while the higher temperature scaling exponent remains unexplained. Our proposed defect-driven MCD hysteresis is consistent with the impurity-induced variable-range hopping observed in low temperature transport measurements.

We expect the bulk defect density to remain relatively uniform throughout the crystal. Polishing the crystal exposes a previously unprobed region of the bulk crystal. A dominant bulk contribution contradicts our observation that spontaneous MCD disappears after a new region of the bulk crystal was exposed by polishing. This interpretation is consistent with transport measurements on thin films reported in Ref. \cite{zhou2026surfacestatedrivenanomaloushalleffect}, which reported a surface-state-driven AHE. Surface defects and impurity-induced variable-range hopping may alter the electronic structure in such a way that causes hysteresis to disappear after polishing. Defects are localized on the atomic length scale, on the order of several \AA, at least one order of magnitude smaller than both our laser spot size and optical penetration depth. Our measurements probe the collective effect of all defects within the illuminated volume and cannot resolve the nature of individual defects or quantitatively estimate the defect density. Our measurements cannot distinguish whether the disappearance of spontaneous MCD results is from change's in the surface electronic structure caused by disorder-induced variable range hopping or a different defect density. If defects are the microscopic origin of the observed hysteresis loop, then they are unlikely to also cause the observed memory effect. The origin of the hysteresis sign change around 200 K observed here and in prior studies \cite{GonzalezBetancourt_AHE2023, Chilcote_stoichometry2024, liu2026strainAHE, smolenski2025strainberrycurvature, bey2024unexpectedtuninganomaloushall, zhou2026surfacestatedrivenanomaloushalleffect} remains unclear. This behavior provides further evidence that the AHE may not be intrinsic.

In conclusion, we have observed spontaneous circular dichroism in a MnTe bulk crystal. The spontaneous MCD, an optical analogue of the AHE, emerges around 250 K, below the reported N\'eel temperature and changes sign around 200 K, consistent with prior AHE results. We also found that the hysteresis loop is shifted away from zero field. We propose the shift arises from a dilute secondary spin population. The direction of the shift depends on the sign of the cooling field. Both the polished bulk crystal and MBE-grown thin film showed no hysteresis, suggesting that the spontaneous MCD in MnTe is not from bulk altermagnetic order. We instead attributed the spontaneous MCD to a defect-driven surface state. More work is needed to understand the nature of the surface defects and the hysteresis sign change around 200 K. Looking forward, we hope our work motivates the community to identify new systems where the AHE is intrinsic to the altermagnetic ground state.  \\

B.M. and L.W. were sponsored by the Army Research Office under Grant No. W911NF-20-2-0166 and W911NF-25-2-0016 for the measurement and data analysis. L.W. also acknowledges support from the Office of Naval Research under Grant No. N00014-24-1-2064  and from Kaufman Foundation under the award number KA2025-151542 for manuscript writing. The single crystal growth work at Rice is supported by the U.S. DOE BES DE-SC0026179 (P.D., E.M.). Part of the materials characterization work at Rice is supported by the Robert A. Welch Foundation under Grant No. C-1839 (P.D.) and C-2114 (A.J. and E.M.). Z.-J.Y and C.-Z.C were supported by the Office of Naval Research under Grant No. N000142412133. The single-crystal growth work by S.H.Y. and Z.M. was performed at the Pennsylvania State University Two-Dimensional Crystal Consortium—Materials Innovation Platform (2DCC-MIP), which is supported by NSF Cooperative Agreement No. DMR-2039351. Film growth at Oak Ridge by M.C. and M.B. was supported by the U. S. Department of Energy (DOE), Office of Science, Basic Energy Sciences (BES), Materials Sciences and Engineering Division (thin film growth). J.W.F.V. acknowledges support from the U.S. Department of Energy under Early Career Award No. DE-SC0025632.

\bibliography{references.bib}
\end{document}